\documentclass[prl,twocolumn,showpacs,preprintnumbers,amsmath,amssymb,times]{revtex4}
\usepackage{graphicx}% Include figure files
\usepackage{dcolumn}% Align table columns on decimal point
\usepackage{bm}% bold math
\usepackage{psfrag}
\usepackage{epsfig}
\usepackage{amsmath}
\usepackage{amssymb}
\usepackage{color}
\usepackage{mathbbol}

\newcommand{\nn}{\nonumber}
\newcommand{\bra}{\langle}

\newcommand{\ket}{\rangle}

\begin{document}

\title{Steady state entanglement in the mechanical vibrations of two dielectric membranes}

\author{Michael J. Hartmann and Martin B. Plenio}
\affiliation{Institute for Mathematical Sciences, Imperial College London, and\\
QOLS, Blackett Laboratory, Imperial College London,
SW7 2BW, United Kingdom}
\email{m.hartmann@imperial.ac.uk}

\date{\today}

\begin{abstract}
We consider two dielectric membranes suspended inside a 
Fabry-Perot-cavity, which are cooled to a steady state 
via a drive by suitable classical lasers. We show that 
the vibrations of the membranes can be entangled in this 
steady state. They thus form two mechanical, macroscopic 
degrees of freedom that share steady state entanglement.
\end{abstract}

%Classification Scheme.
\maketitle
%\tableofcontents

% ---------------------------------------------------------------------------
%
\paragraph*{Introduction --}

Optomechanical systems in which electromagnetic degrees of freedom 
couple to the mechanical motion of mesoscopic or even macroscopic 
objects are promising candidates for studying the transition of a 
macroscopic degree of freedom from the classical to the quantum 
regime. These systems can also be of considerable technological 
use, e.g. for improved displacement measurements \cite{RBM+04} 
and their application in the detection of gravitational 
waves \cite{BV02}.

Optomechanical devices have therefore attracted considerable 
attention in recent years and micromirrors have been cooled by 
radiation-pressure \cite{ACB+06}. In many setups, one of the end
mirrors of a Fabry-Perot cavity undergoes a mechanical vibration
and the coupling between cavity photons and the mirror motion
emerges because the resonance 
frequency of the cavity depends on its length and hence 
on the position of the mirror. Recently devices have been introduced 
in which the motion of a membrane that is inserted into a 
Fabry-Perot cavity formed by rigid mirrors couples to the
cavity mode \cite{TZJ+08,BM08}.

Whereas the ground state and hence the quantum
regime has not yet been reached in experiments, this has 
been predicted to be achievable if the mechanical oscillation 
frequency is larger than the cavity linewidth \cite{WNZK07}, 
a regime that has recently been observed \cite{SRA+08}.
In the quantum regime, it is then interesting to 
explore entanglement in mechanical i.e. macroscopic 
degrees of freedom \cite{ABS02,EPBH04,TZ04}. Possibilities 
to entangle the motion of a cavity micromirror 
with the electromagnetic field in the cavity have thus been 
explored in various approaches \cite{MGVT02,PDV05,VMT07,VGF+07}.

\begin{figure}
\centering
\includegraphics[width=7cm]{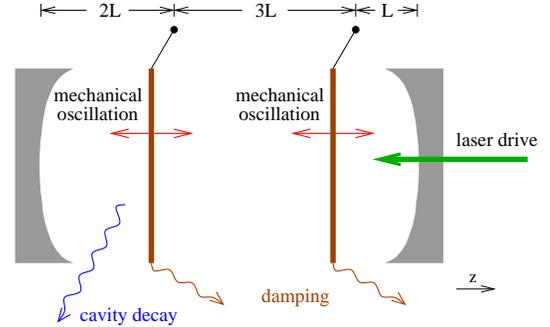}
\caption{\label{setup} The setup: Two mechanically vibrating 
membranes (brown) are suspended inside a Fabry-Perot-cavity, 
which is driven by external lasers (green). Dissipation occurs 
via mechanical damping (brown) and cavity decay (blue).}
\end{figure}

Here, we consider a Fabry-Perot cavity with two dielectric 
membranes suspended in its interior (c.f. figure \ref{setup}) 
and assume that two cavity resonances are driven by external 
lasers. With suitable lasers the mechanical vibrations of 
the membranes are cooled and asymptotically driven into a 
steady state. We show that the mechanical vibrations of 
the two membranes can be entangled in this asymptotic state.
Entanglement between mechanical oscillators has been discussed
previously but was either only found in the transient regime \cite{MGVT02}
(not the steady state) or required either a drive with non-classical light \cite{PDV05}
or mechanical oscillators that had been pre-cooled to very low temperatures \cite{VMT07}.
In contrast, our scheme generates steady state entanglement between
mechanical degrees of freedom by cooling them via radiation
pressure which only uses classical light sources.
Our approach is not restricted to the specific setup mentioned here but also applies
to other devices with optomechanical couplings \cite{ACB+06},
between two mechanical and two cavity modes.

\paragraph*{Model --}

We consider a Fabry-Perot cavity with two dielectric membranes 
in its interior (c.f. figure \ref{setup}).
In this setup the optical resonance frequencies of the cavity
depend on the positions of the membranes and can be derived from the boundary 
conditions of the field in the cavity \cite{SL72}. Let $-3L$ ($3L$) 
and $q_1$ ($q_2$) be the positions of the left (right) rigid mirror 
and the left (right) membrane and $T$ the transmissivity of the membranes.
We denote the field modes in the left, center and right part of the cavity by
$u_1$, $u_2$ and $u_3$. For a mode with wavenumber $k$,
the boundary conditions read \cite{SL72},
$u_1(-3L) = u_3(3L) = 0$, $u_1(q_1 - 0) =  u_2(q_1 + 0)$,
$u_2(q_2 - 0) =  u_3(q_2 + 0)$,
$\frac{d \, u_1(q_1 - 0)}{dz} - \frac{d \, u_2(q_1 + 0)}{dz}
= k \eta u_1(q_1)$ and
$\frac{d \, u_2(q_2 - 0)}{dz} - \frac{d \, u_3(q_2 + 0)}{dz}
= k \eta u_2(q_2)$,
where $\eta = 2 \sqrt{(1-T)/T}$. In analogy to \cite{BM08}, we make the ansatz $u_1 = A \sin(k(q + 3L))$,
$u_2 = B \cos(k q) + \tilde{B} \sin(k q)$ and $u_3 = C \sin(k(q - 3L))$ and
obtain the transcendental equation,
$[ \cos(3 k L) - \eta \cos(k q_1) \sin(k q_1 + 3 k L)]$
$[ \sin(3 k L) + \eta \sin(k q_2) \sin(k q_2 - 3 k L)]$ $+$
$[ \cos(3 k L) + \eta \cos(k q_2)\sin(k q_2 - 3 k L)]$
$[ \sin(3 k L) + \eta \sin(k q_1) \sin(k q_1 + 3 k L)] = 0$,
from which the optical resonance frequencies $\omega = k c$ ($c$ is the speed of light) can be found.

To obtain optomechancial coupling between two optical and two independent mechanical modes (see below), we choose the equilibrium positions of the membranes, $q_{01}$ and $q_{02}$, to be
$q_{01} = - L$ and $q_{02} = 2 L$. For non-vibrating membranes, the optical resonance frequencies of the membrane cavity systems are then given by
$\omega_{an} = \frac{n \pi c}{L}$,
$\omega_{bn} = \frac{n \pi c}{L} - \frac{\theta c}{2 L} + \frac{c}{2 L} \cos^{-1}\left(-\frac{\cos \theta}{2}\right)$,
$\omega_{b'n} = \frac{n \pi c}{L} - \frac{\theta c}{2 L} - \frac{c}{2 L} \cos^{-1}\left(-\frac{\cos \theta}{2}\right)$, and
$\omega_{cm} = \frac{m \pi c}{3 L} + \frac{\pi c}{6 L} - \frac{\theta c}{3 L}$, where $n$ and $m$ are positive integers. For membranes with low transmissivity, the frequencies $\omega_{an}$, $\omega_{bn}$ and $\omega_{cm}$ with $m = 3 n$ lie close together whereas $\omega_{b'n}$ is separated from this triplet and we thus focus on $\omega_{an}$, $\omega_{bn}$ and $\omega_{cm}$.

In the case of vibrating membranes, the optical resonances depend on the
motion of the membranes and their frequencies become functions of $q_1$ and $q_2$,
e.g. $\omega_{an}(q_1,q_2)$
(The assumption that the optical resonances only depend on the membrane positions,
not their momenta, is only valid if the membrane oscillations are much slower than the optical round-trip time, i.e. $\omega_m \ll |\omega_x - \omega_y|$ for $x,y = an, bn, cm$ \cite{Law95},
which we confirm below.).
To obtain these functions, we write them as a power series up to linear order in the membrane positions,
$\omega_x (q_1, q_2) = \omega_x (q_{01}, q_{02}) + \xi_{x1} (q_1 - q_{01}) + \xi_{x2} (q_2 - q_{02})$ for $x = an,bn,cm$ ($q_{01} = -L$ and $q_{02} = 2 L$), expand the transcendental equation
up to linear order in $q_j - q_{0j}$ ($j = 1,2$) and
solve it for zeroth and linear order separately to
obtain $\omega_x (q_{01}, q_{02}) = \omega_x, \xi_{x1}$ and $\xi_{x2}$ for $x = an,bn,cm$.
For our choice of the membrane rest positions, $q_{01} = -L$ and $q_{02} = 2 L$, the
mode $\omega_{an}$ does not couple to the membrane motions, $\xi_{a1} = \xi_{a2} = 0$,
and we discard it.
The other couplings are
$\xi_{b1} \approx \left(\frac{1}{10}+\frac{3}{400} T \right) \frac{n \pi c}{L^2}$,
$\xi_{b2} \approx \left(\frac{2}{5}-\frac{39}{200} T \right) \frac{n \pi c}{L^2}$
and $\xi_{c1} = - \xi_{c2} \approx - \left(\frac{4}{45}-\frac{28}{675} T \right) \frac{m \pi c}{L^2}$ for $T \ll 1$ and $n,m \gg 1$.
Higher order terms in the expansion of the frequencies give rise to additional coupling terms,
also for $\omega_{an}$, but these are negligible compared to the linear couplings. For symmetric membrane rest positions ($q_{01} = - L$, $q_{02} = L$), one would get $\xi_{b1} + \xi_{b2} = \xi_{c1} + \xi_{c2} = 0$ and
the photons would only couple to the breathing mode, $q_1 - q_2$, whereas the center of mass mode,  $q_1 + q_2$, would not be cooled. Note also the two optical modes are needed to cool two mechanical modes.

The corresponding Hamiltonian that describes the motion of the membranes and the
cavity modes reads
\begin{align} \label{ham}
H = & \frac{\omega_m}{2} \sum_{j=1,2} \left(p_j^2 + q_j^2 \right)
+ \sum_{x = bn,cm} \biggl(\frac{\Omega_x}{2} a_x + \text{h.c.} \biggr) \nn \\
+ & \sum_{x = bn,cm} \biggl( \Delta_x + \sum_{j=1,2} \xi_{xj} q_j \biggr) a_x^{\dagger} a_x \, ,
\end{align}
where $p_j$ and $q_j$ are the momentum and position of membrane $j$ ($j = 1,2$).
Both membranes have the same effective mass $m$ and mechanical resonance
frequency $\omega_m$ and the optical modes with creation(annihilation) operators $a_{bn}^{\dagger}(a_{bn})$ and $a_{cm}^{\dagger}(a_{cm})$ are driven by classical lasers with Rabi frequencies $\Omega_{bn}$ and $\Omega_{cm}$.
We have redefined the position variables $q_j - q_{0j} \rightarrow q_j$ and write
the optical modes in frames that rotate at the frequencies of their respective driving lasers, $\Delta_x = \omega_{x0} - \omega_{x,\text{Laser}}$ \cite{GZbook}.
In eq. (\ref{ham}), we have also assumed that each laser only
drives one cavity mode, which sets an upper bound to the permissible Rabi frequencies,
$|\Omega_{bn}|, |\Omega_{cm}| \ll |\omega_{bn} - \omega_{cm}| \approx \frac{5}{12} c \sqrt{T} / L$
(to leading order in $T \ll 1$).
This in turn limits the amount of entanglement that can be generated.

The linear optomechanical couplings, $\xi_{xj} q_j$ can be exploited to cool
the membranes and drive them into a steady state. They will furthermore generate entanglement between the mechanical vibrations via the optical modes as we will show.

\paragraph*{Equations of motion --}
\label{sec:EOM}
Taking into account cavity decay and mechanical damping of the membranes, the Hamiltonian (\ref{ham}) gives rise to the Langevin equations \cite{GZbook},
\begin{align} \label{langevin1}
\dot{a}_x = & -i \biggl( \Delta_x + \sum_{j=1,2} \xi_{xj} q_j - i \frac{\Gamma_x}{2} \biggr) a_x - i \frac{\Omega_x^{\star}}{2} + \sqrt{\Gamma_x} a_x^{\text{in}}\, , \nn \\
\dot{q}_j = & \omega_m p_j \: , \:
\dot{p}_j = - \omega_m q_j -\frac{\gamma}{2} p_j - \sum_{x} \xi_{xj} a_x^{\dagger} a_x + \zeta_j, 
\end{align}
where dots denote time derivatives and $[\cdot]^{\star}$ a complex conjugate. $a_x^{\text{in}}$ and $\zeta_j$ are the optical and mechanical input noises and $\Gamma_x$ and $\gamma$ cavity decay and mechanical damping rates. The relevant nonzero correlation functions of the noise operators are
$\bra a_x^{\text{in}} (t) \left( a_y^{\textrm{in}} \right)^{\dagger} (t') \ket = \delta_{xy} \delta(t-t')$ 
for $x,y = bn,cm$ and 
$\bra \zeta_j (t) \zeta_l (t') \ket = \frac{\gamma}{2} \left( 2 n_{\omega_m} + 1 \right) \delta_{jl} \delta(t-t')$
for $j,l = 1,2$,
where $n_{\omega_m} = \left( \exp(\hbar \omega_m / k_B T ) - 1 \right)^{-1}$ is the thermal phonon number of the mechanical environment at temperature $T$, $k_B$ is Boltzmann's constant and we have assumed $k_B T \gg \hbar \omega_m$.

We split the operators in (\ref{langevin1}) into their steady state 
expectation values and quantum fluctuations, $a_x = c_x + \delta_x$, 
$p_j = P_j + \delta p_j$ and $q_j = Q_j + \delta q_j$. The constant 
steady state expectation values are given by the equations
\begin{equation} \label{steadystate}
\frac{\Omega_x^{\star}}{2} = - \left(\mu_x - i \frac{\Gamma_x}{2} \right) c_x
 \: \, \text{and} \: \:
Q_j = - \sum_x \frac{\xi_{xj}}{\omega_m} |c_x|^2 ,
\end{equation}
where $\mu_x = \Delta_x + \xi_{x1} Q_1 + \xi_{x2} Q_2$.

We are interested in a regime of high photon numbers in the cavity, 
in which the steady state expectation values are much larger 
than the quantum fluctuations. In this regime we can neglect 
all terms of higher than linear order in the fluctuations,
$\delta_x$, $\delta p_j$ and $\delta q_j$ in (\ref{langevin1})
(We have confirmed this approximation numerically.).
The asymptotic state of the quantum fluctuations for the linearized 
equations is then a zero mean Gaussian state which is fully 
characterized by its covariance matrix
$V_{ij} = 2 \textrm{Re} \left\bra \left(O_i - \left\bra O_i \right\ket \right)
\left(O_j  - \left\bra O_j \right\ket \right)\right\ket$,
where $O = (\delta q_1, \delta p_1, \delta q_2, \delta p_2,X_{bn},Y_{bn},X_{cm},Y_{cm})$ with $X_x = (\delta_x + \delta_x^{\dagger})/\sqrt{2}$ and $Y_x = -i (\delta_x - \delta_x^{\dagger})/\sqrt{2}$.
We solve the linearized Langevin equations for the fluctuations to obtain
the steady state covariance matrix $V$ in the same way as in \cite{VGF+07}.
From $V$, the steady state entanglement 
as measured by the logarithmic negativity $E_N$ ($E_N \not= 0$ means the state is entangled) can then be computed \cite{PV06}.

\paragraph*{Steady state entanglement --}

We consider an example where both membranes have a transmissivity $T = 0.2$, an effective mass of $m = 10^{-9}$g and a mechanical resonance frequency of $\omega_m = 10^6$Hz \cite{TZJ+08}.
The cavity is 6mm long, hence $L = 1$mm. For driving lasers of about 1000nm wavelength, the closest cavity modes have numbers $n = 2 \times 10^3$  and $m = 3 n = 6 \times 10^3$.
For these parameters, the optomechanical couplings attain the values $\xi_{bn,1} = 1.90$kHz,
$\xi_{bn,2} = 6.75$kHz and $\xi_{cm,1} = - \xi_{cm,2} = - 4.53$kHz (Note that we work
in units, where $\delta q_1$ and $\delta q_2$ are dimensionless and given in multiples of
$\sqrt{\hbar/(m \omega_m)}$).

Cooling to the quantum mechanical regime is possible if the mechanical 
oscillation frequency is larger than the optical linewidth 
\cite{WNZK07,SRA+08} and we thus assume $\Gamma_{bn} = \Gamma_{cm} 
= \omega_m / 10$. The mechanical $Q$ is taken to be $Q = 10^7$,
consistent with \cite{TZJ+08}. For the mechanical environment, we consider two temperature values,
${\cal T} =8$mK ($n_{\omega_m} = 1000$) and ${\cal T} = 100$mK ($n_{\omega_m} = 13085$).

\begin{figure}
\centering
\includegraphics[width=4cm]{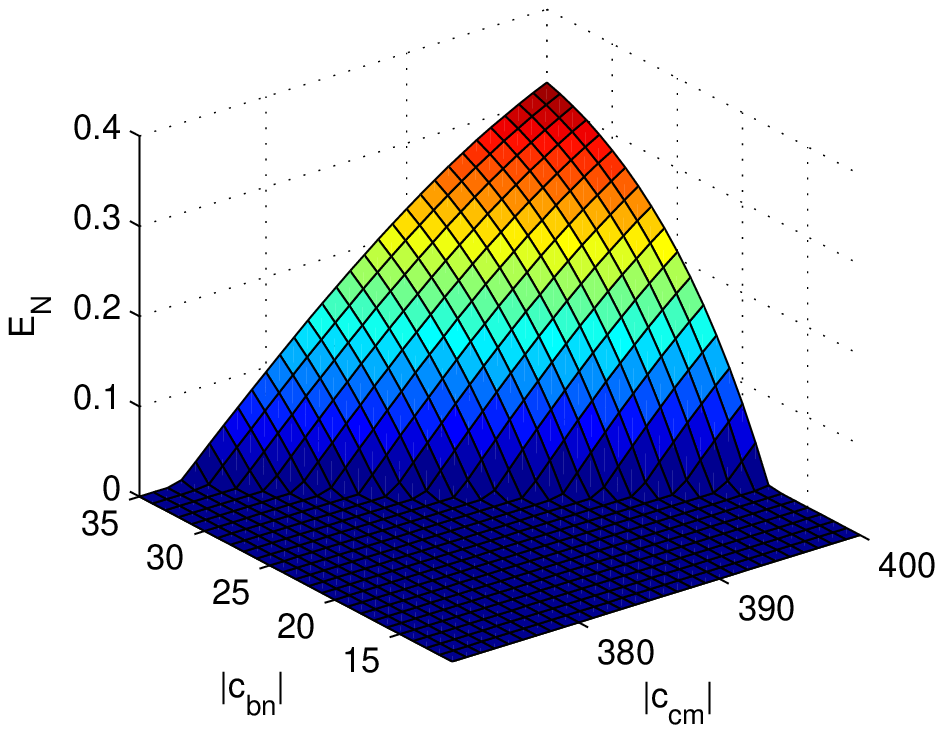}
\includegraphics[width=4cm]{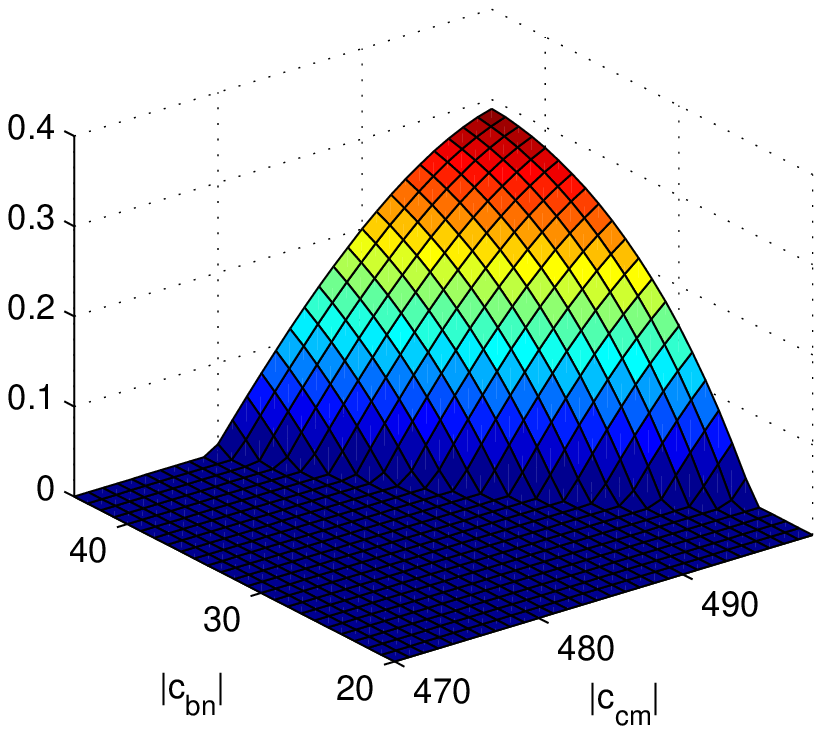}
\caption{\label{entangled1ab} The entanglement of the two mechanical 
vibrations in the steady state as measured by the logarithmic negativity as a function of $c_{bn}$ and $c_{cm}$. The phases of $c_{bn}$ and $c_{cm}$
do not affect the entanglement. Left plot: $\Delta_{bn} = 4.07$MHz, $\Delta_{cm} = 20.84$MHz and
${\cal T} =8$mK ($n_{\omega_m} = 1000$). Right plot:  $\Delta_{bn} = 6.12$MHz, $\Delta_{cm} = 33.18$MHz and ${\cal T} = 100$mK ($n_{\omega_m} = 13085$). The remaining parameters are $\omega_m = 1$MHz, $m = 10^{-9}$g, $T = 0.2$, $L = 1$mm, $n = 2 \times 10^3$, $ = 3 n = 6 \times 10^3$, $q_{01} = -L$, $q_{02} = 2 L$, $\Gamma_{bn} = \Gamma_{cm} = \omega_m / 10$, and $Q = 10^7$ for both plots. $\Delta_{bn}$, $\Delta_{cm}$, $c_{bn}$ and $c_{cm}$ have been optimized numerically for each case.}
\end{figure}
\begin{figure}
\centering
\includegraphics[width=8cm]{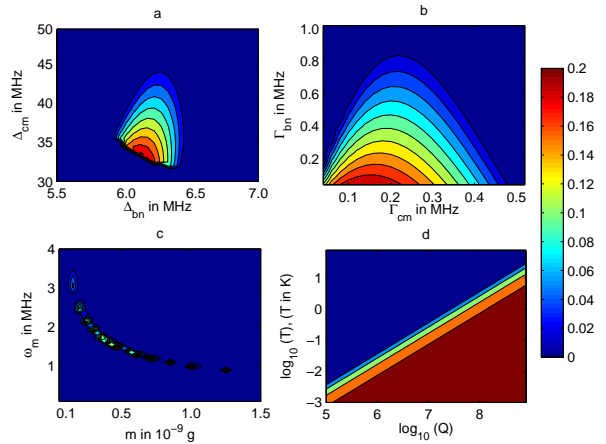}
\caption{\label{entangled1ccdd} The steady state entanglement of the two mechanical vibrations for ${\cal T} = 100$mK, $c_{bn} = 60$ and $c_{cm} = 486$. a: $E_N$ as a function of $\Delta_{bn}$ and $\Delta_{cm}$, b: $E_N$ as a function of $\Gamma_{bn}$ and $\Gamma_{cm}$, c: $E_N$ as a function of $m$ and $\omega_m$ and d: $E_N$ as a function of $Q$ and $T$. All other parameters are the same as in figure \ref{entangled1ab}.}
\end{figure}

Figure \ref{entangled1ab} shows the entanglement of the two mechanical 
vibrations in the steady state measured by the logarithmic negativity, 
$E_N$, as a function of the steady state electromagnetic fields in the 
cavity, $c_{bn}$ and $c_{cm}$. Since the Hamiltonian (\ref{ham}) only contains
the photon numbers, the entanglement is insensitive to the phases of fields in the
cavity and thus also to the phases of $c_{bn}$ and $c_{cm}$.
%$\Delta_{bn} = 4.07$MHz and $\Delta_{cm} = 20.84$MHz (these values have been optimized).
The linearization of eq. (\ref{langevin1}) requires $|c_{bn}|, |c_{cm}| \gg 1$.
In the left plot, we have $|c_{bn}| \sim 30$, but we also find $E_N = 0.195$ for $|c_{bn}| = 60$,
$|c_{cm}| = 386.4$, $\Delta_{bn} = 4.2$MHz and $\Delta_{cm} = 20.9$MHz.
The values for $Q$ and $T$, we assume here, are currently hard to achieve simultaneously but
the entanglement persists in a larger parameter range as shown in figure \ref{entangled1ccdd}d.

The Rabi frequencies $\Omega_{bn}$ and $\Omega_{cm}$ that are
needed to generate the values of $c_{bn}$ and $c_{cm}$ in figure \ref{entangled1ab} are less
than 11GHz (left plot) and 23GHz (right plot). The difference between the resonance frequencies $\omega_{bn}$ and $\omega_{cm}$
on the other hand is
$|\omega_{bn} - \omega_{cm}| \approx 57$GHz and the separation of these two modes from
other resonances is much larger, so that the lasers indeed only drive one resonance mode
as assumed in eq. (\ref{ham}). $|\omega_{bn} - \omega_{cm}| \propto L^{-1}$ so that $|\omega_{bn} - \omega_{cm}|$ would even be larger for a shorter cavity. By reducing the cavity length one could thus employ stronger driving lasers and create substancial entanglement even at higher ${\cal T}$.
The input laser powers $P_x$ are related to the Rabi
frequencies by $P_x = \hbar \omega_{x,\text{Laser}} |\Omega_x|^2 / (4 \Gamma_x)$, which
implies that laser powers between 0.6 pW and 60 $\mu$W are required.
Furthermore $\omega_m \ll |\omega_{bn} - \omega_{cm}|$ and the derivation of the cavity resonances
and consequently the form of Hamiltonian (\ref{ham}) are well justified.

%
% \begin{figure}
% \centering
% \includegraphics[width=8cm]{nnent1.eps}
% \caption{\label{nnent} Characteristics of the steady state for ${\cal T} = 8$mK. {\bf a:} phonon number $n_1$ of membrane 1, {\bf b:} phonon number $n_2$ of membrane 2, {\bf c:} largest real part of the eigenvalues of the matrix A and {\bf d:} the entropy of the reduced density matrix of the vibrations. All parameters as in figure \ref{entangled1ab}.}
% \end{figure}
%
The steady states are furthermore characterized by phonon numbers of the vibration fluctuations, 
$n_j = \frac{1}{2} \left( \delta p_j^2 + \delta q_j^2 - 1 \right)$, of 
$n_1 \le 3, n_2 \le 5$ for ${\cal T} = 8$mK and $n_1 \le 5, n_2 \le 10$ for ${\cal T} = 100$mK,
and an entropy of the reduced density matrix of the vibrations, 
$S_m = - \text{Tr}_{\text{photons}}(\rho \log_2 \rho)$ of less than 4, $S_m \le 4$, for both,
${\cal T} = 8$mK and ${\cal T} = 100$mK.
The steady state is thus indeed in the quantum regime. The linearized Langevin equations 
can be cast in matrix form $\dot{O} = O A + n$, where $n$ is the 
vector of the noises \cite{VGF+07}. We have confirmed that all eigenvalues of $A$
have negative real parts below 1kHz which ensures that there is a unique steady state
that is reached within milliseconds.

To further corroborate the robustness of the entanglement, we studied its dependence on fluctuations in the driving lasers and on variations in several system parameters.
The results for ${\cal T} = 100$mK are shown in fig. \ref{entangled1ccdd}.
Fig. \ref{entangled1ccdd}a shows the dependence of $E_N$ on the detunings of driving lasers, $\Delta_{bn}$ and $\Delta_{cm}$ ($c_{bn} = 60$ and $c_{cm} = 486$),
fig. \ref{entangled1ccdd}b the dependence on the cavity decay rates $\Gamma_{bn}$ and $\Gamma_{cm}$, fig. \ref{entangled1ccdd}c the dependence on the membranes resonance frequency, $\omega_m$ and its effective mass, $m$ and fig. \ref{entangled1ccdd}d the dependence on the mechanical $Q$ and the
environmen temperature ${\cal T}$.
In all cases there is a substantial parameter region which shows entanglement. Note that we assign $E_N = 0$ to all points where there is no well defined steady state due to heating, i.e. where an eigenvalue of $A$ has a positive real part. Furthermore, our driving fields are optimized for the values in figure \ref{entangled1ab}. For different values of $\Gamma_{bn}$, $\Gamma_{cm}$, $Q$, $\gamma$, $\omega_m$ or $m$, slightly modified laser drives will yield more entanglement.

One can obtain some intuitive indications why the membrane vibrations become entangled.
The Hamiltonian corresponding to the linearized Langevin equations for the fluctuations $\delta p_1, \delta q_1, \delta p_2, \delta q_2, \delta_{bn}$ and $\delta_{cm}$ is
$H = \frac{\omega_m}{2} \sum_j \left( \delta p_j^2 + \delta q_j^2 \right)
+ \sum_x \mu_x \delta_x^{\dagger} \delta_x
+ \sum_{j,x} \left( \xi_{xj} c_x \delta q_j \delta_x^{\dagger} + \text{h.c.} \right)$.
In the parameter regime of interest, we have $\mu_x \gg |\xi_{xj} c_x|$ and the photon degrees of freedom can be adiabatically eliminated to obtain the effective Hamiltonian
$\mathcal{H} = \frac{\omega_m}{2} ( \delta p_1^2 + \delta p_2^2)
+ \frac{\omega_m + \nu_1}{2} \delta q_1^2 + \frac{\omega_m + \nu_2}{2} \delta q_2^2
+ \frac{\nu_{12}}{2} \delta q_1 \delta q_2$,
where $\nu_j = - 2 \sum_x \xi_{xj}^2 |c_x|^2 / \mu_x$ and
$\nu_{12} = - 4 \sum_x \xi_{x1} \xi_{x2} |c_x|^2 / \mu_x$.
The ground state of $\mathcal{H}$ and hence states close to it are entangled in
regimes where $|\nu_{12}| \sim \omega_m$ or larger, which is the case for the parameters in figure \ref{entangled1ab}.

\paragraph*{Entanglement verification --}

To verify the created entanglement quantitatively in an experiment, 
several quadrature correlations need to be measured \cite{DGCZ00,EPBH04}.
This may be achieved by 
employing at least two further weak probe lasers, similar in spirit to the scheme in \cite{VGF+07}, which drive cavity modes that do not participate in the entanglement generation, 
i.e. modes with $n \not= 2 \times 10^3$ or $m \not= 6 \times 10^3$. 
The equation of motion for the fluctuations of the probe field, $\delta_y$, reads $\dot{\delta_y} = - i \mu_y \delta_y + \sqrt{\Gamma_y} a_y^{\text{in}} - i \frac{c_y}{\sqrt{2}} (\xi_{y1} C_1 + \xi_{y2} C_2)$ in a frame that rotates at the frequency of the probe laser, $\omega_L$. Here $\delta q_j = \frac{1}{\sqrt{2}} (C_j + C_j^{\dagger})$ and we have assumed $|\xi_j c_y | \ll \omega_m$ and applied a rotating wave approximation. In a Fourier transformed picture in the laboratory frame
this equation reads $-i (\omega - \mu_y) \delta_y(\omega) = \sqrt{\Gamma_y} a_y^{\text{in}}(\omega) - i \frac{c_y}{\sqrt{2}} [\xi_{y1} C_1(\omega - \omega_L) + \xi_{y2} C_2(\omega - \omega_L)]$.
Applying standard input-output formalism \cite{GZbook}, 
$a_y^{\text{out}} (\omega) = \sqrt{\Gamma_y} [c_y \delta(\omega - \omega_L) + \delta_y(\omega)] - [c_y^{\text{in}} \delta(\omega - \omega_L) + a_y^{\text{in}}]$,
the output field is given by $a_y^{\text{out}} (\omega) = -\frac{\omega - \mu_y - i \Gamma_y}{\omega - \mu_y} a_y^{\text{in}}(\omega) +
(\sqrt{\Gamma_y} c_y - c_y^{\text{in}}) \delta (\omega - \omega_L) +
\frac{c_y}{\omega-\mu_y} \sqrt{\frac{\Gamma_y}{2}} (\xi_{y1} C_1 (\omega - \omega_L) +
\xi_{y2} C_2 (\omega - \omega_L))$, where $c_y^{\text{in}}$ is the input field of the probe laser. The background terms $c_y$ and $c_y^{\text{in}}$ only contribute for $\omega = \omega_L$. Homodyne measurements on the output field thus allow to measure $\xi_{y1} C_1 (\omega - \omega_L) +
\xi_{y2} C_2 (\omega - \omega_L)$. The second probe laser on a mode with different $\xi_{y1}$ and $\xi_{y2}$ measures a different linear combination of $C_1$ and $C_2$ and hence  a different quadrature of the membrane vibrations. As the steady state of the membranes allows for repeated measurements, two probe lasers enable a reconstruction of the covariance matrix $V$, where the precision is limited by the input noise, $a_y^{\text{in}}$.

\paragraph*{Conclusions --}

The scheme presented here allows to generate steady state entanglement 
of the motion of two dielectric membranes, which are suspended inside 
a Fabry-Perot cavity with a cavity decay rate that is lower than the 
mechanical resonance frequency of the membranes. The scheme only 
requires a drive by classical light and can work for environment temperatures
up to a few Kelvin. With increasing environment temperatures stronger driving
lasers and therefore shorter cavities with larger mode seperation are needed.
The output fields of weak probe lasers that do not perturb the entanglement generation can be used to measure quadratures of the vibrational modes and thus enable
a verification of the created steady state entanglement.

MJH wants to thank Ignacio Wilson-Rae for fruitful discussions and helpful comments. This work is part of the QIP-IRC supported by EPSRC (GR/S82176/0), the EU Integrated
Project Qubit Applications (QAP) supported by the IST directorate
as Contract Number 015848' and was supported by the EPSRC grant EP/E058256 and
the Alexander von Humboldt foundation.

\end{document}